\documentclass[reqno,keywordsasfootnote]{article}
\usepackage[english]{babel}
\usepackage[latin1]{inputenc}
\usepackage{amsmath,amsthm,amsfonts,amssymb,times}
\numberwithin{equation}{section}
\usepackage[final]{epsfig}

\def\sig{\underline{\sigma}}
\newcommand{\state}[1]{\ensuremath{\left\langle #1 \right \rangle}}

\newtheorem{thm}{Theorem}

\def\F{\mathcal{F}}
\def\H{\mathcal{H}}

\def\Sys{QRFIM}
\newcommand{\lop}[1]{\ensuremath{\kappa_{#1}}}

\begin{document}  

\title{\vspace{-3cm} 
\LARGE On Spin Systems with Quenched Randomness: \\ Classical and Quantum}

\author{ Rafael L. Greenblatt$^{(a)}$
\footnote{Work supported by NSF Grant DMR-044-2066 and AFOSR Grant AF-FA9550-04}  
    \quad Michael Aizenman$^{(b)}$  
 \footnote{Work supported by NSF Grant DMS-060-2360}   \quad    Joel L. Lebowitz$^{(c) *}$
 \\[2ex]   
{\normalsize $^ {(a)}$  Department of Physics and Astronomy}\\  {\normalsize  Rutgers University, Piscataway NJ 08854-8019, USA}\\[1ex] 
{\normalsize  $^{(b)}$ Departments of Physics and Mathematics}  \\  
{\normalsize  Princeton University,  Princeton NJ 08544, USA  } \\[1ex]   
{\normalsize $^ {(a)}$  Departments of Mathematics and Physics}\\  
{\normalsize  Rutgers University, Piscataway NJ 08854-8019, USA} 
  }



\date{December 5, 2009}
\maketitle

\begin{flushright}
\emph{Dedicated to Nihat Berker }\mbox{ \hspace{1.2 cm}}\\ 
\emph{on the occasion of his 60th birthday} \\[1ex] 
\end{flushright}

\begin{abstract}
The rounding of first order phase transitions by quenched randomness is stated in a form which is applicable to both classical and quantum systems:  The free energy, as well as the ground state energy, of a spin system on a $d$-dimensional lattice is continuously differentiable with respect to any parameter in the Hamiltonian to which some randomness has been added when $d \leq 2$.   This implies absence of jumps in the associated order parameter, e.g., the magnetization in case of a random magnetic field.
A similar result applies in cases of continuous symmetry breaking for $d \leq 4$.
Some questions concerning the behavior of related order parameters in such random systems are discussed.
\end{abstract}

\section{Introduction}

The effect of quenched randomness on the equilibrium and transport properties of macroscopic systems is a subject of great theoretical and practical interest  which is close to Nihat's heart.  He and his collaborators~\cite{HuiBerker.PRL,Berker.PRB.90,MBHFP} made important contributions
to the study of the changes brought about by such randomness in phase transitions occurring in the pure (non-random) system.  These effects can be profound in low dimensions.  Their   understanding evolved in a somewhat interesting way.
As was argued by Imry, Ma and others~\cite{HuiBerker.PRL,YM,AYM}, even a very small amount of randomness can change the nature of the transition in dimensions $d\le 2$.  The validity of the reasoning involved was contentious as it conflicted with the prediction of a dimensional-reduction principle, which was arrived at through perturbative calculations carried to all orders~\cite{PS}.
Rigorous work by Imbrie~\cite{Imbrie.PRL} and Bricmont and Kupiainen~\cite{BK.PRL} showed that, in the discrete case, the rounding phenomenon does not extend beyond two dimensions, as given by the Imry-Ma argument.
Aizenman and Wehr~\cite{AW.PRL,AW} provided a rigorous proof of the rounding effect in low dimensions, which although limited to classical spin systems included a large variety of types of randomness not previously considered.  As we have recently shown~\cite{QIMLetter}, with some modification this proof can be extended to quantum systems.  Here we present these results in a unified way, and discuss related issues which appear in specific examples of interest.

To formulate the results in a general, but not the most general way, we consider a lattice system of
generalized spin variables
%
$\sig=\{\sigma_x\}$, $x \in \mathbb{Z}^d$, where $\sigma_x$ is an operator in a finite dimensional Hilbert space or, in the classical case, a function taking values in a finite set.  The Hamiltonian is of the form
\begin{equation}
\label{genHam}
\mathcal{H}_\epsilon = \mathcal{K}(\sig;\underline{J})-h\sum_x \lop{x} - \epsilon \sum_x \eta_x \lop{x} = \mathcal{H}_0 - \epsilon \sum_x \eta_x \lop{x}
\end{equation}
where $\underline{J}$ denotes a set of interaction parameters,  $\{\lop{x}\}$ are translates of some local operator $\lop{0}$,
 and $h$ and $\epsilon$ are real parameters.  The quenched disorder is represented by $\{\eta_x\}$, a family of independent, identically distributed random variables with an absolutely continuous distribution with mean zero and finite third moment (weaker conditions are possible, but more awkward to state).  $\mathcal{K}(\sig;\underline{J})$ may be translation invariant and nonrandom, or it can include additional random terms but we will not discuss the latter case.
It is assumed here that the interactions decay faster than $(\mbox{distance})^{-\gamma}$, with $\gamma > 3d/2$ for the case described below in theorem~\ref{bigThm} and $\gamma > d+2$ for theorem~\ref{bigThm2}.

We refer to the $\eta$s as random fields, although in general they may also be associated with some other parameters, e.g.  random bond strengths as in the models studied in~\cite{HuiBerker.PRL,MBHFP}.

An example of a system of this type (with $\lop{x}=\sigma_x^{(3)}$) is the ferromagnetic transverse-field Ising model with a random longitudinal field~\cite{Senthil} (henceforth \Sys), with
 \begin{equation} \label{thisHam}
\mathcal{H}_\epsilon = -\sum J_{x-y} \sigma_x^{(3)} \sigma_y^{(3)} - \sum \left[ \lambda \, \sigma_x^{(1)} + (h+ \epsilon \eta_x) \, \sigma_x^{(3)} \right] \end{equation} where $\sigma_x^{(i)}$ $(i=1,2,3)$ are single-site Pauli matrices, and $J_{x-y} \geq 0$.  When $\lambda=0$ this is the classical Ising system with a random field.

We will examine phase transitions where the order parameter is the volume average of the expectation value of $\lop{x}$ with respect to an equilibrium state.     These states are the infinite volume limits of sequences of Gibbs states in finite boxes $\Lambda \subset \mathbb{Z}^d$ with some boundary conditions, for which the expectation values are denoted here by $\state{ - } _\Lambda$.

As is well known,
this order parameter is related to the directional derivatives $(\pm)$ of the (quenched) free energy density given by
\begin{equation}
\F(T, \underline{J}, h, \epsilon)\ =\  -T \lim_{\Lambda \nearrow \mathbb{Z}^d} \frac{1}{|\Lambda|} \log Z_\Lambda,
\end{equation}
where $Z_\Lambda$ is the partition function of the system on a finite domain $\Lambda$ of size $|\Lambda|$.  Under fairly modest conditions~\cite{AW,vuillermot1977tqr}, it is known that this limit exists, is independent of the sequence of boundary conditions and (almost certainly) of $\eta$.  Furthermore, it is a concave function of $h$, which means it has directional derivatives; and these are related to the order parameter by
\begin{equation}\label{direc}
-\frac{\partial \F}{\partial h-} \leq \lim_{\Lambda \nearrow \mathbb{Z}^d} \frac{1}{|\Lambda|} \sum_{x \in \Lambda} \state{\lop{x}}_\Lambda \leq -\frac{\partial \F}{\partial h+}
\end{equation}
for any sequence of sufficiently regular (e.g. rectangular) domains $\Lambda$  converging to $\mathbb{Z}^d$.

More can be said, in particular in the classical case.  There, convexity arguments imply that if the mean free energy density $\F(T,h)$ is differentiable in $h$, i.e.
\begin{equation} \label{eq:diff}
\frac{\partial \F}{\partial h-}  = \frac{\partial \F}{\partial h+} \equiv \frac{\partial \F}{\partial h}  \, ,
\end{equation} then the block averages of $\lop{x}$ over
large rectangular volumes $B$
do not asymptotically deviate from the thermodynamic value $-\frac{\partial \F}{\partial h}$.  More explicitly, there is a positive function $g(\delta)$ such that for every $\delta >0$:
\begin{equation}
\lim_{\Lambda   \to \mathbb{Z}^d}
\rm{Av}\left(  \rm{Prob}_\Lambda \left(  \left|\frac{1}{|B|} \sum_{x\in B} \lop{x}  + \frac{\partial \F}{\partial h} \right|  > \delta  \right)\,  \right)  \ \le \  A\, e^{-g(\delta) |B|}
\end{equation}
for some $A<\infty$, where $\rm{Av}$ is an average over the random fields, and $\rm{Prob}_\Lambda \left( - \right)$ is  the probability with respect to the Gibbs equilibrium measures for any sequence of  $\Lambda$'s which converges to $\mathbb{Z}^d$.   This general principle, which follows from the convexity of the free energy, can be used to show that the differentiability of the `quenched' free energy density, i.e., \eqref{eq:diff} which is the subject of our discussion, implies the vanishing of the `Short Long Range Order parameter' as well as of the `Long Long Range Order Parameter' which appears in~\eqref{direc}.

Combining the results known for some time for classical systems (where $\H_0$ and $\lop{x}$ are functions of a local configuration)~\cite{AW}, with those recently proven also for quantum spin systems (where they are self adjoint operators)~\cite{QIMLetter}, we obtain:
\begin{thm}\label{bigThm}
In dimensions $d\le 2$, any system of the form of \eqref{genHam}
has $\frac{\partial \F}{\partial h -}=\frac{\partial \F}{\partial h +}$ for all $h$, and $T\ge 0$, provided $\epsilon \neq 0$.    \end{thm}

In view of equation~\eqref{direc}, or the above comment, this means that no first-order (i.e. discontinuous) transition with $\lop{x}$ as the order parameter is possible in the presence of randomness in the local value of $h$: any such transition present in the system described by $\H_0$ is ``rounded''.


For systems where this spontaneous ordering would break a continuous symmetry, the results can be extended to $d \leq 4$.  Here we must have observables which transform as vectors, so we replace~\eqref{genHam} with
\begin{equation}
\mathcal{H}_\epsilon =  \mathcal{K}  - \sum (\vec{h} + \epsilon \vec{\eta}_x) \cdot \vec{\sigma}_x = \mathcal{H}_0 - \epsilon \sum \vec{\eta}_x \cdot \vec{\sigma}_x, \label{ONHam} \end{equation}
where
%
%
$\vec{h}$, $\vec{\eta}_x$, and $\vec{\sigma}_x$ are quantities which transform as vectors in a space of dimension $N\ge 2$ and $\mathcal{K}$ is invariant under such rotations.

\begin{thm}\label{bigThm2}
For the systems described above, with the random fields $\vec{\eta}_x$ having a rotation-invariant distribution, the free energy
is continuously differentiable in $\vec{h}$ at $\vec{h}=0$ whenever $\epsilon \neq 0$, $d \leq 4$.
\end{thm}
As before, the proof for classical systems is found in~\cite{AW} and the proof for quantum systems in~\cite{QIMLetter}.

\section{Higher Order Odd Correlations}

It can be proven rigorously for the models described by~\eqref{thisHam} with purely ferromagnetic interactions ($J_{x-y}\geq 0$) and $\lambda=0$
that in the pure ($\epsilon=0$) system, not only the magnetization $\state{\sigma^{(3)}_x}$ but all odd correlations, e.g. $\state{\sigma^{(3)}_x \sigma^{(3)}_y \sigma^{(3)}_z}$, $x \neq y \neq z$, are discontinuous at $h=0$ for $T<T_c$~\cite{lebowitz1977coexistence}.  The same is presumably true quite generally, particularly for the QRFIM with $\lambda\neq 0$; a discontinuity in the magnetization brings with it also a discontinuity in all odd correlation functions.  We may now ask whether the randomness in the external field, corresponding to $\epsilon \neq 0$, which removes the discontinuity in $\state{\sigma_x}$ also removes the discontinuity in all the odd correlation functions when averaged with respect to the randomness.

For the cases covered by~\eqref{thisHam} when $J_{x-y} \geq 0$, this question can be answered in the affirmative, including for $\lambda \neq 0$.  In this case the system satisfies the Fortuin-Kasteleyn-Ginibre (FKG) inequalities~\cite{FKG}.  Using these, it is possible to show that whenever all Gibbs states agree on the magnetization per site $\state{\sigma_x}$ (i.e. when the spontaneous magnetization $M$ is zero) they also agree on all other local observables (the argument for random systems is spelled out in~\cite{bovier2006smd}).  We do not have any such result for the case when $J_{x-y}$ in~\eqref{thisHam} contains also some negative terms.  We expect of course that this will be the case more generally, but we lack a proof.

\begin{figure}
\begin{center}
\includegraphics[width=0.6\textwidth]{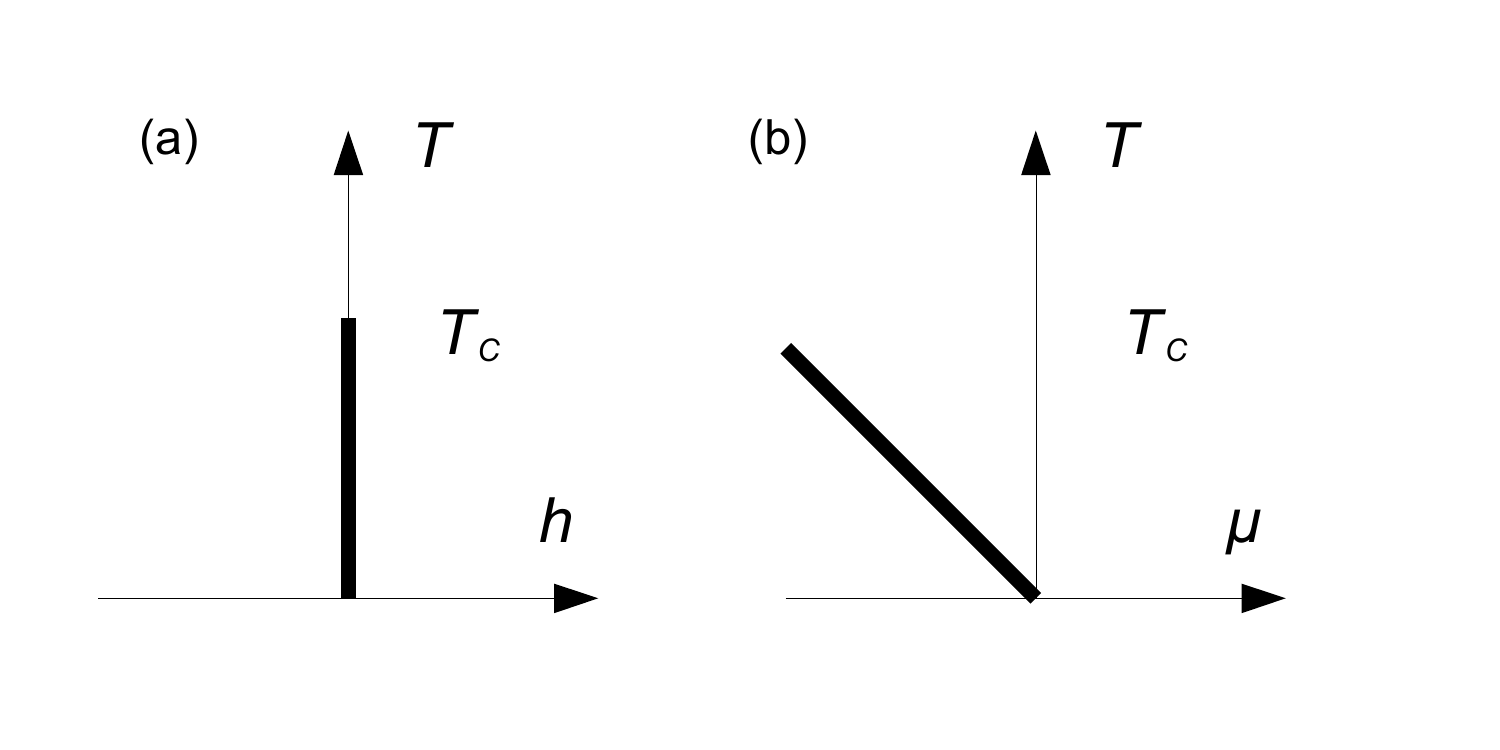}
\end{center}

\caption{Phase diagrams of (a) the simple Ising model and (b) Ising lattice gas in two dimensions; thick lines are first order phase transitions.}
\label{IsingFig}
\end{figure}
For an instructive example let us consider the phase diagram of the Ising model~\eqref{thisHam} with $\lambda=0$ and $d=2$ with nearest-neighbor interactions $J_{\hat{e}}=J>0$ ($\hat{e} \in \{(0,1),(1,0)\}$), $\epsilon=0$ (see Figure~\ref{IsingFig}a).  The free energy is analytic everywhere except on the bold part of the $T$ axis.  Along that line the free energy is continuous but its derivatives are discontinuous in all directions transversal to the $T$ axis.  That is, the free energy surface has a seam, i.e. a line of `kink singularities'.



Those singularities will be rounded through the addition of arbitrarily small randomness to the field $h$.  However, the kink singularity would not be eliminated by small randomness in the bond strengths.  That is, with $J$ replaced by $J+\epsilon b_{x,\hat{e}}$ with the $b_{x,\hat{e}}$ i.i.d random variables,
there would still be spontaneous magnetization at low temperatures, though   $T_c$ may change, and so will possibly the nature of the singularity in $\F$ at $T_c$.
The randomness is tangential to the seam, corresponding to ``orthogonal randomness''  in the  terminology of~\cite{Berker.PRB.90}.

Let us now consider this system in terms of lattice gas variables.  Setting $\xi_x = \tfrac{1}{2} (1+\sigma_x) \in \{0,1\}$, the nonrandom Hamiltonian now takes the form
\begin{equation}\label{lgHam}
\mathcal{H}_0  \ =\   -4 J \sum_{<x,y>} \xi_x \xi_y - \mu \sum_x \xi_x + \mbox{constant}.
\end{equation}
The chemical potential $\mu$ is related to the parameters of the spin model by $\mu=2h-8J$, and so the phase diagram is now of the form shown in Figure~\ref{IsingFig}b.  At the bold line the free energy has a discontinuous derivative in both variables, and randomness in either $J$ or $\mu$ will eliminate this first order transition, rounding the seam in $\F(T,\mu,0)$.  To avoid destroying the first order transition we have to add randomness to $J$ and $\mu$ in a correlated manner, so as to make it `tangential' to the seam, i.e., in the form done in \eqref{thisHam} in the $\sigma$-variables.

Similar reasoning can be applied quite generally.  In particular if we consider the free energy $\F(T,\underline{J},h,\epsilon=0)$ for the system described by~\eqref{thisHam} with $\underline{J}$ containing interaction parameters multiplying both even and odd functions of the spins then, for $T<T_c$, $\F$ will have ``multidimensional seams'' in directions transverse to all the odd terms -- which is why they are all discontinuous in the absence of randomness.  Adding then randomness to any odd term, e.g. to the one spin term as in~\eqref{thisHam}, then rounds the discontinuity in all of them.
If on the contrary we add noise only to the even terms, this is tangential to the ``seams'' and doesn't destroy the first order transition.

This is also the case when adding noise in the \Sys  ~\eqref{thisHam}    to the transverse component of the field, i.e. letting $\lambda\sum \sigma^{(1)}_x \to \sum (\lambda + \epsilon \eta_x) \sigma^{(1)}_x$.  As proven in~\cite{campanino1991lgs} this does not destroy the discontinuity in $\state{\sigma^{(3)}_x}$.

The situation is similar in the case when the spin at site $x$, $S_x$, assumes the values $S_x=-1,0,1$.  A particular version of such a system is the Blume-Capel model,
whose nonrandom version is defined by
\begin{equation}
\mathcal{H}_0  \ =\  -J \sum_{<x,y>} (S_x - S_y)^2 - \Delta \sum S_x^2 - h \sum S_x.
\end{equation}
\begin{figure}
\begin{center}
\includegraphics[width=0.6\textwidth]{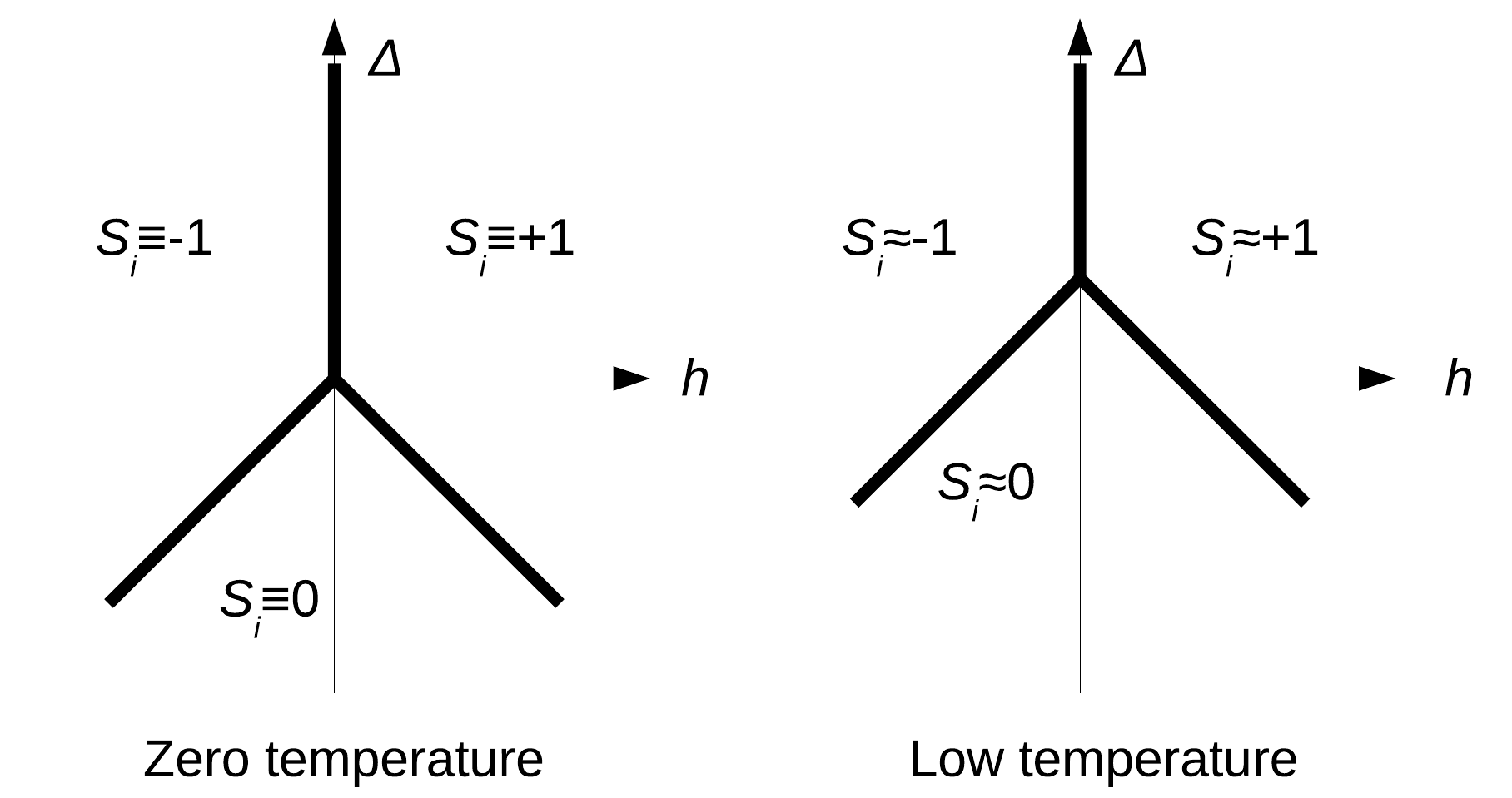}
\end{center}
\label{BEGFig}
\caption{Sections of the phase diagram for the Blume-Capel model.}
\end{figure}
The low-temperature phase diagram of this system is known rigorously from Pirogov-Sinai theory~\cite{PS1,PS2}.  A section is shown in Figure~\ref{BEGFig} alongside the corresponding diagram at zero temperature.

In analogy with the Ising case, randomizing the magnetic field $h$, by $h\mapsto h+\epsilon \eta_x$, will destroy  all the first order lines in $d\le 2$ (in $d=1$ this would be relevant only at $T=0$).
Randomness in  $J$ or $\Delta$ on the other hand should not destroy the discontinuity along the $h=0$ line.  This is in accord with the analysis in \cite{HuiBerker.PRL}.
Similar results will hold for the quantum version of this model with $S_x$ now interpreted as $S^{(3)}_x$, and a transverse field term $\sum_x S_x^{(1)}$ added.


\section{Effects of bond randomness on the discontinuity of spontaneous magnetization}

While the `tangential disorder', in the sense discussed above,  will not eliminate the discontinuity in the order parameter associated with the transversal field, e.g. the spontaneous magnetization at low enough temperatures, it \emph{may}   eliminate the possible discontinuity of the spontaneous magnetization if one was present at $T_c$ in the non-random model.
A general statement to that effect  was proposed in~\cite{HuiBerker.PRL}.

 An relevant example is provided by   the
 two dimensional Potts models with nearest neighbor ferromagnetic interactions.
At   $Q>4$ these models exhibit a  discontinuity in the spontaneous magnetization at $T_c$, which  coincides with a jump in the energy density.
 While Theorem 1 assures that randomization of the bond strengths will eliminate the energy discontinuity, it does not address the persistence or not of the discontinuity in the spontaneous magnetization.   In ~\cite{HuiBerker.PRL} it was argued that also the latter discontinuity  will be eliminated, but this remains an interesting open challenge from the rigorous point of view~\cite{CLM}.

%

%

%

\section{One dimension, inverse square interaction}

Another class of models exhibiting a jump in the spontaneous magnetization are
the one-dimensional $Q$-state Potts models, at $Q\ge 1$, with  ferromagnetic pair interactions $J_{x-y}$ which asymptotically behave as $J^+/|x-y|^2$.  It is known that such systems have a phase transition, and that these exhibit somewhat unusual features~\cite{ACCN,ImbrieNewman}(and references therein).   Among those is the discontinuity of the spontaneous ``magnetization'' $M(T)$ at zero ``magnetic'' field, which jumps at $T_c$ from zero to a value satisfying  $M^2 J^+ / T_c \geq 1$.

At the moment it is not clear to us  whether for the $1/{r^2}$  Potts models the
discontinuity in $M(T)$ may coincide with a discontinuity in the energy.  That will of course correspond to a jump  at $T_c$ in the derivative of $\F$ with respect to $T$.   If that occurs then  the system will have $T=T_c$ at least $Q+1$ phases.  In the absence of  energy discontinuity one may expect that at the critical temperature the systems will have  $Q$ ordered phases in analogy with the results of ~\cite{lebowitz1977coexistence} for Ising systems.
%

%


Let us consider now the effects of the additions of a random symmetry - breaking ``magnetic''  field and of randomization, at suitably decaying rates, of the spin-spin interaction terms.  As mentioned in the Introduction, the proof of rounding phenomenon~\cite{AW} applies also to the long range $d=1$ models as long as the pair interaction decays faster than $|x-y|^{-3/2+\delta}$ (this ensures that the interaction energy between an interval of size $L$ and the rest of the system is less than $L^{1/2}$).  Thus, in the presence of  random magnetic field  the spontaneous magnetization $M(T)$ will vanish for all $T$.  Randomization of the pair interaction would remove the discontinuity in the energy, if there was one, but  unlike for the case discussed above we do not expect bond randomness to remove the discontinuity in $M(T)$ at the (possibly altered) critical temperature.
%
The reason for the possible difference with the two dimensional models is the bound
$M^2 J^+ / T \geq 1$ which relates the discontinuity in the $d=1$ models to the purely long range part of the interaction.

\section*{Acknowledgements}
It has been both instructive and a personal pleasure for us to know Nihat Berker.    We (MA and JLL) thank him for the many interesting discussions of physics as well as for his kind hospitality, which we enjoyed on different occasion at MIT and in Istanbul.   We wish you Nihat  many fruitful and enjoyable years past this birthday.


\end{document}